\documentclass[graybox]{svmult}

\usepackage{mathptmx}       
\usepackage{helvet}         
\usepackage{courier}        
\usepackage{type1cm}        
\usepackage{makeidx}         
\usepackage{graphicx}        
\usepackage{multicol}        
\usepackage[bottom]{footmisc}

\makeindex             

\begin{document}

\title*{Quadruple-peaked Line-of-sight Velocity Distributions in Shell Galaxies}
\titlerunning{Quadruple-peaked LOSVD in Shell Galaxies}
\author{Ivana Ebrov\'{a}, Lucie J\'{i}lkov\'{a}, Bruno Jungwiert, Kate\v{r}ina Barto\v{s}kov\'{a}, Miroslav K\v{r}\'{i}\v{z}ek, Tereza Bart\'{a}kov\'{a}, and Ivana Stoklasov\'{a}}
\authorrunning{Ebrov\'{a} et al.}
\institute{Ivana Ebrov\'{a}, Bruno Jungwiert, and Miroslav K\v{r}\'{i}\v{z}ek 
\at Astronomical Institute, Academy of Sciences of the Czech Republic; Faculty of Mathematics and Physics, Charles University in Prague, \email{ivana@ig.cas.cz}
\and Lucie J\'{i}lkov\'{a} 
\at ESO Santiago, Chile; Faculty of Science, Masaryk University, Brno, Czech Republic
\and Kate\v{r}ina Barto\v{s}kov\'{a} and Tereza Bart\'{a}kov\'{a} 
\at Faculty of Science, Masaryk University, Brno, Czech Republic
\and Ivana Stoklasov\'{a} \at
Astronomical Institute, Academy of Sciences of the Czech Republic
}

\maketitle

\vskip-1.2truein

\abstract{We present an improved study of the expected shape of the line-of-sight velocity distribution in shell galaxies. We found a simple analytical expression connecting prominent and in principle observable characteristics of the line profile and mass-distribution of the galaxy. The prediction was compared with the results from a test-particle simulation of a radial merger.}

\section*{Quadruple-peaked Spectral-line Profile}
\label{sec1:losvd}
Stellar shells are observed in almost half of elliptical and S0 galaxies that live in a low galactic density environment, see e.g.~\cite{col01}. They are thought to be by-products of galaxy mergers~\cite{quinn84}. The most regular shell systems are believed to result from a nearly minor radial merger in which the satellite galaxy is dissolved by tidal forces and its stars begin to oscillate in the potential of the host galaxy at close-to-radial orbits. The stars accumulate at their turning points and create shells. 

The shape of line-of-sight velocity distribution (LOSVD) in the vicinity of the shell edge for a stationary shell was studied by~\cite{mk98}. They predicted a double-peaked spectral-line profile and proposed to use spectroscopy to probe the dark matter distribution of a galaxy that contains shells using the profiles of stellar absorption lines.

Nevertheless, shells are not stationary features: stars of the satellite galaxy have a continuous energy distribution, and therefore the shell edge is, at different times, made of stars of different energies, as they continue to arrive at their respective turning points. Thus, the shell front moves outwards from the center of the host galaxy with its velocity given by the mass distribution of the host galaxy. Therefore, both of the original double peaks in the spectral line are split into two, resulting in a \textit{quadruple-peaked shape} \cite{ji10}. 
Taking the shell's velocity and the cumulative mass of the host galaxy to be constant near the edge of the shell, we found an approximate analytical description for the positions of the peaks in the LOSVD (for details see \cite{eb11}).

\begin{figure}[t]
%
\begin{center}
\includegraphics[angle=270,scale=.3]{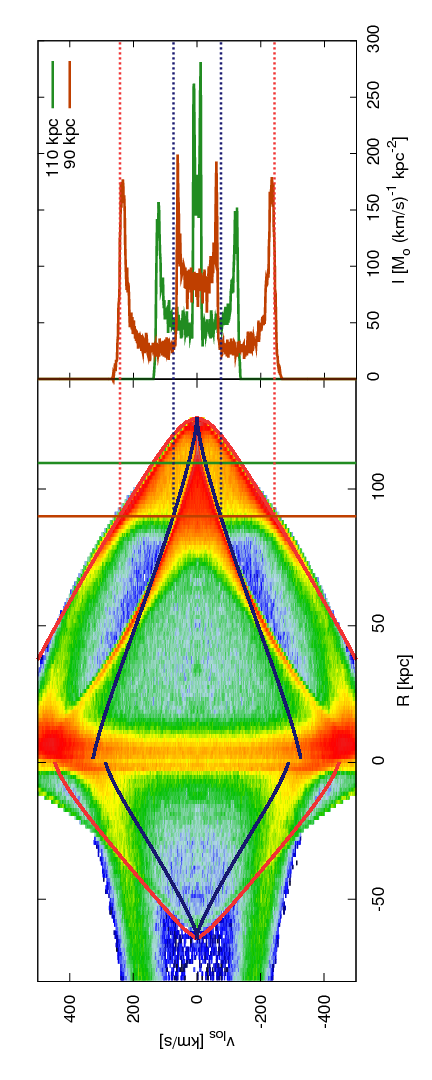}
\end{center}
\caption{\textit{left:} LOSVD map of the simulated shell galaxy (only stars of the satellite galaxy are taken into account). The two apexes of the wedges seen in the map (at zero velocity) correspond to the two shells, the black curves show the velocity maxima position obtained from our approximation. \textit{right:} LOSVDs (cuts of the map shown in the left plot) of stars belonging to the right shell at two different galactocentric distances---90 and 110 kpc---red and green profiles respectively. Dashed lines show the locations of the maxima in our approximation for the red (90 kpc) profile.}
\label{fig:ebrova21}       
\end{figure}

To study the LOSVD more in detail we carried out a test-particle simulation of a radial merger of dwarf (dE) and giant elliptical (gE) galaxies, leading to a formation of shells. The potential of the gE galaxy is represented with a luminous de Vaucouleurs sphere and an NFW dark halo. See Fig.~\ref{fig:ebrova21} for comparison of LOSVD from simulation and the analytical approximation. If the velocity maxima were measured, the approximation could be used to constrain the mass distribution of the host galaxy.

\begin{acknowledgement}
We acknowledge the support by the grant No. 205/08/H005 (Czech Science
Foundation), the project SVV 261301 (Charles University in Prague), EAS
grant covering registration fee at JENAM 2010, the grant MUNI/A/0968/2009
(Masaryk University in Brno), the grant LC06014, Center for Theoretical
Astrophysics (Czech Ministry of Education) and research plan AV0Z10030501
(Academy of Sciences of the Czech Republic).
\end{acknowledgement}
%


%


\begin{thebibliography}{99.}%
%
%
\bibitem{col01}
Colbert, J. W. et al., 2001, AJ, 121, 808

\bibitem{eb11}
Ebrov\'{a}, I. et al., 2011, in preparation

\bibitem{ji10}
J\'{i}lkov\'{a}, L. et al., 2010, ASCP, 423, 243

\bibitem{mk98}
Merrifield, M.\,R. \& Kuijken, K., 1998, MNRAS, 297, 1292

\bibitem{quinn84}
Quinn, P.\,J., 1984, ApJ, 279, 596

\end{thebibliography}
\end{document}